# Ultrathin compound semiconductor on insulator layers for high performance nanoscale transistors

Hyunhyub Ko[1,2,3,†], Kuniharu Takei[1,2,3,†], Rehan Kapadia[1,2,3,†], Steven Chuang[1,2,3], Hui Fang[1,2,3], Paul W. Leu[1,2,3], Kartik Ganapathi[1], Elena Plis[5], Ha Sul Kim[5], Szu-Ying Chen[4], Morten Madsen[1,2,3], Alexandra C. Ford[1,2,3], Yu-Lun Chueh[4], Sanjay Krishna[5], Sayeef Salahuddin[1], Ali Javey[1,2,3,*]

[1]Electrical Engineering and Computer Sciences, University of California, Berkeley, CA, 94720.

[2]Materials Sciences Division, Lawrence Berkeley National Laboratory, Berkeley, CA 94720.

[3]Berkeley Sensor and Actuator Center, University of California, Berkeley, CA, 94720.

[4]Materials Science and Engineering, National Tsing Hua University, Hsinchu 30013, Taiwan, R. O. C.

[5]Center for High Technology Materials, University of New Mexico, Albuquerque, NM 87106.

[†] These authors contributed equally to this work.

* Correspondence should be addressed to A.J. (ajavey@eecs.berkeley.edu).

Over the past several years, the inherent scaling limitations of electron devices have fueled the exploration of high carrier mobility semiconductors as a Si replacement to further enhance the device performance[1,2,3,4,5,6,7,8]. In particular, compound semiconductors heterogeneously integrated on Si substrates have been actively studied[7,9,10], combining the high mobility of III-V semiconductors and the well-established, low cost processing of Si technology. This integration, however, presents significant challenges. Conventionally, heteroepitaxial growth of complex multilayers on Si has been explored[9,11,12,13]. Besides complexity, high defect densities and junction leakage currents present limitations in the approach. Motivated by this challenge, here we utilize an epitaxial transfer method for the integration of ultrathin layers of single-crystalline InAs on Si/SiO$_2$ substrates. As a parallel to silicon-on-insulator (SOI) technology[14], we use the abbreviation "XOI" to represent our compound semiconductor-on-insulator platform. Through experiments and simulation, the electrical properties of InAs XOI transistors are explored, elucidating the critical role of quantum confinement in the transport properties of ultrathin XOI layers. Importantly, a high quality InAs/dielectric interface is obtained by the use of a novel thermally grown interfacial InAsO$_x$ layer (~1 nm thick). The fabricated FETs exhibit an impressive peak transconductance of ~1.6 mS/μm at $V_{DS}$=0.5V with ON/OFF current ratio of greater than 10,000 and a subthreshold swing of 107-150 mV/decade for a channel length of ~0.5 μm.

Epitaxial lift-off and transfer of crystalline microstructures to various support substrates has been shown to be a versatile technique for applications ranging from optoelectronics to large-area electronics[15,16,17,18]. Specifically, high performance, mechanically flexible macro-electronics and photovoltaics have been demonstrated on plastics, rubbers, and glass substrates by this method[19,20,21]. Here, we use a modified epitaxial transfer scheme for integrating *ultrathin* InAs layers with *nanometer-scale* thicknesses on Si/SiO$_2$ substrates for use as high performance nanoscale transistors. The nanoscale thick InAs layers are fully depleted which is an important criteria for achieving high performance FETs with respectable OFF currents based on small band gap semiconductors. The transfer is achieved without the use of adhesive layers, thereby allowing for purely inorganic interfaces with low interface trap densities and high stability. The process schematic for the fabrication of InAs XOI substrates is shown in Figure 1a (see Methods for the details).

Atomic force microscopy (AFM) was utilized to characterize the surface morphology and uniformity of the fabricated XOI substrates. Figures 1b-c show representative AFM images of an array of InAs NRs (~18 nm thick) on a Si/SiO$_2$ substrate, clearly depicting the smooth surfaces (< 1 nm surface roughness) and high uniformity of the enabled structures over large areas. Uniquely, the process readily enables the heterogeneous integration of different III-V materials and structures on a single substrate through a multi-step epitaxial transfer process. To demonstrate this capability, a two-step transfer process was utilized to form ordered arrays of 18 and 48 nm thick InAs NRs that are perpendicularly oriented on the surface of a Si/SiO$_2$ substrate (Figs. 1d-e). This result demonstrates the potential ability of the proposed XOI technology for generic heterogeneous and/or hierarchical assembly of crystalline semiconducting materials. In the future, a similar scheme may be utilized to enable the fabrication of both *p*- and *n*- type transistors on the same chip for complementary electronics based on the optimal III-V semiconductors.

To shed light on the atomic structure of the interfaces, cross-sectional transmission electron microscopy (TEM) images of an InAs XOI device were taken and are shown in Figure 2. The high-resolution TEM (HRTEM) image (Fig. 2c) illustrates the single-crystalline structure of InAs NRs (~13 nm thick) with atomically abrupt interfaces with the SiO$_2$ and ZrO$_2$ layers. The TEM image of the InAs/SiO$_2$ interface does not exhibit visible voids (Fig. 2c), although only a small fraction of the interface is examined by TEM. As described in more depth later in this paper, InAs NRs were thermally oxidized prior to the top-gate stack deposition to drastically lower the interfacial trap densities. The thermally grown InAsO$_x$ layer is clearly evident in the HRTEM image (Fig. 2c) with a thickness of ~1 nm.

Long-channel, back-gated field-effect transistors (FETs) based on individual NRs were fabricated in order to elucidate the intrinsic electron transport properties of InAs NRs as a function of thickness. The process scheme involved the fabrication of XOI substrates with the desired InAs thickness followed by the formation of source/drain (S/D) metal contacts by lithography and lift-off (~50 nm thick Ni). The p$^+$ Si support substrate was used as the global back-gate with a 50 nm thermal SiO$_2$ as the gate dielectric. Ni contacts were annealed at 225ºC for 5 min in a N$_2$ ambient to enable the formation of low resistance contacts to the conduction band of InAs (Fig. S6) [22]. The transfer characteristics at $V_{DS}$=0.1V of the back-gated XOI FETs with a channel length, $L$~5 µm and InAs thicknesses of 8-48 nm are shown in Figure 3a. Two trends are clearly evident from the measurements. First, the OFF current monotonically increases with increasing thickness due to the reduced electrostatic gate coupling of the back-gate. Second, the ON current increases with InAs thickness due to the thickness dependency of electron mobility, $\mu_n$. Since $L$~ 5µm, the devices are effectively operating in the diffusive regime, thereby, enabling the direct extraction of the field-effect mobility by using the relation $\mu_{n,FE} = (g_m)(L^2/C_{ox}V_{DS})$, where $g_m = dI_{DS}/dV_{GS}\big|_{V_{DS}}$ is the transconductance and $C_{ox}$ is the gate oxide capacitance (Fig. S5). For this analysis, parasitic resistances were ignored since Ni forms near ohmic metal contacts[22]. The gate oxide capacitance was estimated from the parallel plate capacitor model $C_{ox} = (\varepsilon A)/d$, where $\varepsilon$=3.9 and $d$=50 nm are the dielectric constant and thickness of SiO$_2$, respectively. The effect of quantum capacitance, $C_Q$ was neglected due to the relatively thick gate dielectrics used in this study (*i.e.*, $C_{ox}$<<$C_Q$). Figure 3b shows the peak $\mu_{n,FE}$ as a function of InAs thickness, $T_{InAs}$. The mobility at first linearly increases with thickness for $T_{InAs}$ < ~18 nm with a slope of ~221 (cm$^2$/Vs)/nm, beyond which it nearly saturates at $\mu_{n,FE}$ ~5,500

cm$^2$/Vs. The measured XOI field-effect mobility is close to the reported Hall mobilities for InGaAs (~10,000 cm$^2$/Vs)[10] and InAs (13,200 cm$^2$/Vs)[23] quantum well (QW) structures. It should be noted that the Hall mobility is typically higher than the field-effect mobility for any given material as a number of device and surface state contributions to carrier transport are not accounted in the Hall effect measurements.

To shed light on the observed mobility trend, the low-field phonon mobility, $\mu_{n,phonon}$ was calculated as $\mu_{n,phonon} = e/(m^* \langle 1/\tau \rangle)$, where e is the electronic charge and $m^*$ is the effective mass (see Supp. Info.). Average scattering rate $\langle 1/\tau \rangle$ is calculated from

$$\left\langle \frac{1}{\tau} \right\rangle = \frac{\int \frac{1}{\tau(E)} \frac{\partial f_0}{\partial E} dE}{\int \frac{\partial f_0}{\partial E} dE}$$

where $f_0$ is the equilibrium Fermi-Dirac distribution function. $\tau(E)$ was calculated using Fermi's golden rule, with the matrix elements of the scattering potentials evaluated in the basis of the NR eigenfunctions. Both acoustic and optical (including polar) phonon scattering events were considered[24]. The calculated $\mu_{n,phonon}$ vs. $T_{InAs}$ is shown in Fig. 3b. For small thicknesses, the mobility linearly increases with the thickness. This behavior is attributed to the gradual transition of the channel from a 2D to 3D system as the NR thickness is increased, with more transport modes (*i.e.*, sub-bands) contributing to the current flow. As the thickness surpasses the Bohr radius of bulk InAs (~34 nm), the electronic structure of NRs approaches the 3D regime, resulting in a mobility saturation for $T_{InAs}$>~35 nm to the well-known bulk value of InAs (~40,000 cm$^2$/Vs)[25]. While the onset thickness of saturation closely matches the experiments, there is 5-10× discrepancy in the actual mobility values. This is expected since the extracted data represents the field-effect mobility, consisting of phonon scattering along with other device contributions, including interface trap states, surface roughness scattering, and vertical-field-induced mobility degradation. Both surface roughness and vertical-field (i.e., gate field) induce additional carrier scattering events at the surface/interface, while the primary effect of interface trap states is to deteriorate the modulating the channel conductance (i.e., charge density) by the gate-field. These effects degrade the extracted $g_m$ and thereby $\mu_{n,FE}$. To simulate $\mu_{n,FE}$, a full device simulation was performed (see Supp. Info.). An interface trap density, $D_{it}= 6\times10^{12}$ states $cm^{-2}eV^{-1}$ was used as the fitting parameter. The simulated *I-V* characteristics of XOI back-gated FETs are shown in Figure 3a. Clearly, the simulated *I-V* curves match the experimental data closely for all InAs thicknesses, especially in the ON-state. Next, peak $\mu_{n,FE}$ was extracted from simulation and plotted as a function of $T_{InAs}$ (Fig. 3b), once again closely matching the experimental $\mu_{n,FE}$. The close matching of the experimental and simulated results demonstrate the effectiveness of the XOI platform as a clean and predictable material system for exploring high performance devices while highlighting the critical role of quantum confinement and surface contributions on the transport properties of InAs, even at relatively large thicknesses. It should be noted that since the ribbon width used in this work is 10~30 times larger than the thickness, there is minimal dependence of the device performance on the NR width (Fig. S13), thereby, the structures can be effectively treated as thin films.

In order to explore the performance limits of InAs XOI devices, top-gated FETs with high-κ gate dielectrics and *L*~0.5 μm were fabricated. Briefly, Ni S/D contacts were lithographically patterned on InAs NRs followed by the atomic layer deposition of ~8 nm thick $ZrO_2$ ($\varepsilon \sim 20$) as the gate dielectric. A local top-gate (Ni, 50 nm thick), underlapping the S/D electrodes by ~100 nm was then lithographically patterned. Importantly, thermal oxidation of InAs was found to significantly improve the interfacial properties and FET characteristics (Fig.

S8). In this regard, prior to the S/D contact formation, the XOI substrates were first treated with 3% NH$_4$OH to remove the native oxide followed by the thermal oxidation at 350ºC for 1 min to form ~1 nm thick InAsO$_x$ layer as observed from TEM analysis (Fig. 2c). Figure 4a shows a typical $I_{DS}$-$V_{GS}$ characteristic of a top-gated FET, consisting of an individual ~18 nm thick InAs NR with a width of ~320 nm. The XOI FET exhibits a respectable $I_{ON}/I_{OFF}$~10$^4$, a subthreshold swing of $SS$=d$V_{GS}$/d(log$I_{DS}$) ~ 150 mV/decade (Fig. 4a), and a peak $g_m$ ~ 1.6 mS/µm at $V_{DS}$=0.5V (Fig. S9). The lowest measured $SS$ for our XOI FETs is ~107 mV/decade (Fig. S10) as compared to InAs and InGaAs QW-FETs in literature which have exhibited $SS$ ~ 70 and 75 mV/dec, respectively [10,23]. The devices reported here use a relatively thick gate dielectric which can be scaled down in the future to further improve the gate electrostatic control and the $SS$ characteristics. The single NR transistor output characteristic is shown in Fig. 4b, delivering an impressive $I_{ON}$~1.4 mA/µm at an operating voltage, $V_{DD}$=$V_{DS}$=$V_{GS}$=1V. To further analyze the performance, a full device simulation was performed. A close match of the experimental data is obtained with $D_{it}$=10$^{11}$ states cm$^{-2}$eV$^{-1}$ used as the fitting parameter (see Supp. Info., Fig. S7), which is a ~60× improvement over devices without any surface treatment (*i.e.*, with a native oxide layer). The fitted $D_{it}$ values represent only estimation. Note that while *C-V* measurement is conventionally utilized for $D_{it}$ extraction in Si devices, doing so is rather challenging and prone to a large uncertainty for narrow bandgap semiconductors, such as InAs[26]. In the future, the development of more accurate techniques for $D_{it}$ measurement in InAs XOI devices is needed. The explored thermal oxidation process for surface passivation is counter-intuitive as the previous works have focused on the removal of surface oxides[7]. We speculate that unlike the native oxide layer, thermal oxidation results in the formation of a dense oxide with minimal dangling bonds. Similar to the thermally grown $SiO_2$, the thermal oxide of InAs provides an ideal and simple surface passivation layer, addressing one of the important challenges in InAs devices.

In conclusion, a new technology platform and device concept for the integration of ultrathin layers of III-V semiconductors directly on Si substrates is demonstrated, enabling excellent electronic device performances. While in this work we focus on InAs as the active channel material, in the future, other compound semiconductors could be explored using a similar scheme. In the future, research on the scalability of the process for 8" and 12" wafer processing is needed. We envision that wafer bonding of $Si/SiO_2$ and III-V wafers followed by the etch release of the sacrificial layer may be utilized to enable a manufacturable process scheme for ultrathin XOI devices.

**Methods summary**

Single-crystalline InAs thin films (10-100 nm thick) were grown epitaxially on a 60 nm thick $Al_{0.2}Ga_{0.8}Sb$ layer on bulk GaSb substrates (see Supp. Info., Fig. S1). Polymethylmethacrylate (PMMA) patterns with a pitch and line-width of ~840 nm and ~350 nm, respectively, were lithographically patterned on the surface of the source substrate. The InAs layer was then pattern etched into nanoribbons (NRs) by using a mixture of citric acid (1 g/ml of water) and hydrogen peroxide (30%) at 1:20 volume ratio, which was chosen for its high selectivity and low resulting InAs edge roughness[27]. To release the InAs NRs from the source substrate, the AlGaSb layer was selectively etched by ammonium hydroxide (3%, in water) solution for 110 min[28]. Note that the selective etching of the AlGaSb layer was high enough not to affect the nanoscale structure of the InAs NRs (see Supp. Info., Fig. S2). Next, an elastomeric polydimethylsiloxane (PDMS) substrate (~2 mm thick) was used to detach the partially released InAs NRs from the GaSb donor substrates and transfer them onto $Si/SiO_2$ (50 nm, thermally grown) receiver substrates by a stamping process (Fig. S3-S4) [29]. Notably, in this process scheme, the initial epitaxial growth process is used to control the thickness of the transferred InAs NRs, while the lithographically defined PMMA etch mask is used to tune the length and width.

**Figure Captions**

**Figure 1. Ultrathin InAs XOI fabrication scheme and AFM images. a**, Schematic procedure for the assembly of InAs XOI substrates by an epitaxial transfer process. The epitaxially grown, single-crystalline InAs films are patterned with PMMA and wet etched into NR arrays. A subsequent selective wet etch of the underlying AlGaSb layer and the transfer of NRs by using an elastomeric PDMS slab result in the formation of InAs NR arrays on Si/SiO$_2$ substrates. **b-c,** AFM images of InAs NR arrays on a Si/SiO$_2$ substrate. The NRs have a length of ~10 μm, height of 20 nm and width of ~300 nm. **d-e**, AFM images of InAs NR superstructures on a Si/SiO$_2$ substrate, consisting of two layers of perpendicularly oriented NR arrays with 18 and 48 nm thicknesses as assembled by a two-step epitaxial transfer process.

**Figure 2. Cross-sectional TEM analysis of InAs XOI substrates. a**, A TEM image of an array of three InAs NRs on a Si/SiO$_2$ substrate. **b**, A magnified TEM image of an individual ~13 nm thick InAs NR on a Si/SiO$_2$ (~50 nm thick) substrate. The NR is coated with a ZrO$_2$/Ni bilayer (~15 and ~50 nm, respectively) which acts as a top-gate stack for the subsequently fabricated FETs. **c**, A HRTEM image showing the single-crystalline structure of an InAs NR with abrupt atomic interfaces with ZrO$_2$ and SiO$_2$ layers on the top and bottom surfaces, respectively. A ~1 nm thick InAsO$_x$ interfacial layer formed by thermal oxidation and used for surface passivation is clearly evident.

**Figure 3. Back-gated, long-channel InAs XOI FETs. a**, The experimental (solid lines) and simulated (dashed lines) $I_{DS}$-$V_{GS}$ characteristics of back-gated (50 nm SiO$_2$ gate dielectric) XOI FETs at $V_{DS}$=0.1V with $L$~5 μm for different InAs NR thicknesses (8, 13, 18, 48 nm). Each FET

consists of a single NR. **b**, The experimental and simulated peak field-effect electron mobilities of InAs NRs as a function of NR thickness. The calculated phonon mobility is also shown.

**Figure 4. Top-gated InAs XOI FETs. a**, Transfer characteristics of a top-gated InAs XOI FET, consisting of an individual NR (~18 nm thick) with $L$~0.5 μm and 8 nm thick $ZrO_2$ gate dielectric. A device schematic (top) and a representative SEM image (bottom) of a top-gated FET are shown in the inset. **b**, Output characteristics of the same device shown in (**a**). NRs were thermally oxidized at 350ºC for 1 min to form ~1 nm thick interfacial InAsOx layer for surface passivation of InAs.

**Acknowledgements**

This work was funded by MARCO/MSD Focus Center, Intel Corporation, and BSAC. The materials characterization part of this work was partially supported by a LDRD from Lawrence Berkeley National Laboratory. A.J. acknowledges a Sloan Research Fellowship, NSF CAREER Award, and support from the World Class University program at Sunchon National University. R.K. and M.M. acknowledge an NSF Graduate Fellowship and a postdoctoral fellowship from the Danish Research Council for Technology and Production Sciences, respectively. S.K. acknowledges support from AFOSR Contract FA9550-10-1-0113. Y.-L.C. acknowledges support from National Science Council, R.O.C through No. NSC 98-2112-M-007-025-MY3.

**Author Contributions**

H.K., K.T. and A.J. designed the experiments. H.K., K.T., S.C., H.F., E.P., H.-S.K., M.M. and A.C.F. carried out the experiments. R.K. and P.W.L. performed device simulations. K.G. and S.S. performed mobility calculations. S.-Y.C. and Y.-L.C. performed TEM imaging. H.K., K.T., R.K., P.W.L., K.G., S.K., S.S. and A.J. contributed to analyzing the data. H.K., K.T., R.K. and A.J. wrote the paper while all authors provided feedback.

**Additional information**

Supplementary information accompanies this paper online.

**References**

1. Lundstrom, M. Moore's law forever? *Science* **299**, 210–211 (2003).

2. Heyns, M., Tsai, W. Ultimate scaling of CMOS logic devices with Ge and III-V materials. *MRS Bull.* **34**, 485-488 (2009).

3. Theis, T. N., Solomon, P. M. It's time to reinvent the transistor! *Science* **327**, 1600-1601 (2010).

4. Chau, R., Doyle, B., Datta, S., Kavalieros, J., Zhang, K. Integrated nanoelectronics for the future. *Nature Mater*. **6**, 810-812 (2007).

5. Javey, A., Guo, J., Wang, W., Lundstrom, M., Dai, H. Ballistic carbon nanotube transistors. *Nature* **424**, 654-657 (2003).

6. Wong, P. H.-S. Beyond the conventional transistor. *Solid-State Electron.* **49**, 755–762 (2005).

7. Wu, Y. Q., Xu, M., Wang, R. S., Koybasi, O., Ye, P. Y. High performance deep-submicron inversion-mode InGaAs MOSFETs with maximum $G_m$ exceeding 1.1 mS/um: new HBr pretreatment and channel Engineering. *IEEE IEDM Tech. Digest* **2009**, 323-326 (2009).

8. Bryllert, T., Wernersson, L. E., Froberg, L. E., Samuelson, L. Vertical high-mobility wrap-gated InAs nanowire transistor. *IEEE Electron Device Lett.* **27**, 323-325 (2006).

9. Liu, Y. *et al*. Device physics and performance potential of III-V field-effect transistors. *Fundamental of III-V Semiconductor MOSFETs* (Springer, New York, 2010).

10. Radosavljevic, M. *et al.* Advanced high-k gate dielectric for high-performance short-channel $In_{0.7}Ga_{0.3}As$ quantum well field effect transistors on silicon substrate for low power logic applications. *IEEE IEDM Tech. Digest* **2009**, 319-322 (2009).

11. Javorka, P. *et al.* AlGaN/GaN HEMTs on (111) silicon substrates. *IEEE Electron Device Lett.* **23**, 4-6 (2002).

12. Balakrishnan, G. *et al*. Room-temperature optically-pumped GaSb quantum well based VCSEL monolithically grown on Si (100) substrate, *Electron. Lett.* **42**, 350-351 (2006).

13. Yonezu, H. Control of structural defects in group III-V-N alloys grown on Si. *Semicond. Sci. Technol.* **17**, 762-768 (2002).

14. Celler, G.K., Cristoloveanu, S. Frontiers of silicon-on-insulator. *J. Appl. Phys.* **93**, 4955-4978 (2003).

15. Yablonovitch, E., Hwang, D. M., Gmitter, T. J., Florez, L. T. & Harbison, J. P. Van der Waals bonding of GaAs epitaxial liftoff films onto arbitrary substrates. *Appl. Phys. Lett.* **56**, 2419–2421 (1990).

16. Kim, D.-H. *et al*. Ultrathin silicon circuits with strain-isolation layers and mesh layouts for high-performance electronics on fabric, vinyl, leather, and paper. *Adv. Mater.* **21**, 3703–3707 (2009).

17. Melosh, N. *et al*. Ultrahigh density nanowire lattices and circuits. *Science* **300**, 112-115 (2003).

18. Yokoyama, M. *et al*. III-V-semiconductor-on-insulator n-channel metal-insulator-semiconductor field-effect transistors with buried $Al_2O_3$ layers and sulfur passivation: Reduction in carrier scattering at the bottom interface. *Appl. Phys. Lett*. **96**, 142106 (2010).

19. Yuan, H.-C., Ma, Z. Microwave thin-film transistors using Si nanomembranes on flexible polymer substrate. *Appl. Phys. Lett.* **89**, 212105 (2006).

20. Kim, D.-H. *et al*. Stretchable and foldable silicon integrated circuits. *Science* **320**, 507-511 (2008).

21. Yoon, J. *et al*. GaAs photovoltaics and optoelectronics using releasable multilayer epitaxial assemblies. *Nature* **465**, 329-333 (2010).

22. Chueh, Y.-L. *et al*. Formation and characterization of $Ni_x$InAs/InAs nanowire heterostructures by solid source reaction. *Nano Lett*. **8**, 4528–4533 (2008).

23. Kim, D.-H. *et al.* Scalability of Sub-100 nm InAs HEMTs on InP Substrate for Future Logic Applications*. IEEE Trans. Elec. Dev*. **57**, 1504 (2010).

24. Lundstrom, M. Carrier scattering. *Fundamentals of Carrier Transport* (Cambridge University Press, Cambridge, 2000).

25. Mikhailova, M. P. Indium Arsenide. *Handbook Series of Semiconductor Parameters, vol 1: Elementary Semiconductors and A3B5 Compounds Si, Ge C, GaAs, GaP, GaSb InAs, InP, InSb* (World Scientific, 1996).

26. Martens, K. *et al*. On the correct extraction of interface trap density of MOS devices with high-mobility semiconductor substrates. *IEEE Trans. Electron Devices* **55**, 547-556 (2008).

27. DeSalvo, G. C., Kaspi, R., Bozada, C. A. Citric acid etching of $GaAs_{1-x}Sb_x$, $Al_{0.5}Ga_{0.5}Sb$, and InAs for heterostructure device fabrication, *J. Electrochem. Soc*. **141**, 3526-3531 (1994).

28. Yoh, K., Kiyomi, K., Nishida, A., Inoue, M. Indium arsenide quantum wires fabricated by electron beam lithography and wet-chemical etching. *Jpn. J. Appl. Phys*. **31**, 4515-4519 (1992).

29. Meitl, M. A. *et al*. Transfer printing by kinetic control of adhesion to an elastomeric stamp. *Nature Mater*. **5**, 33-38 (2006).

Figure 1

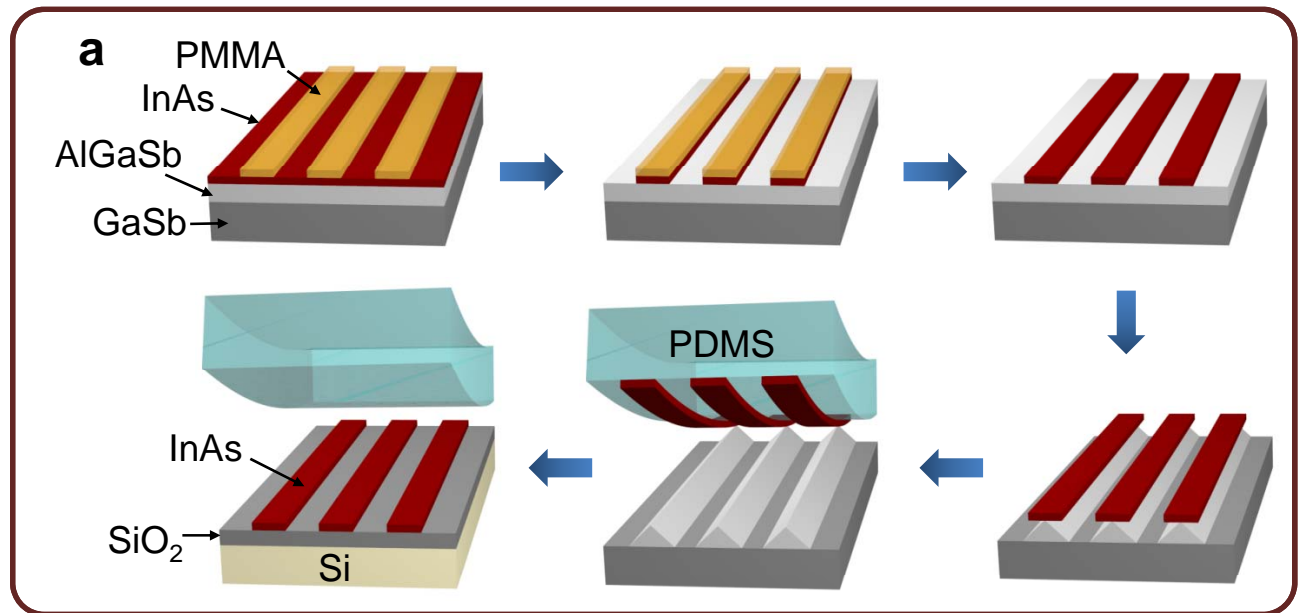
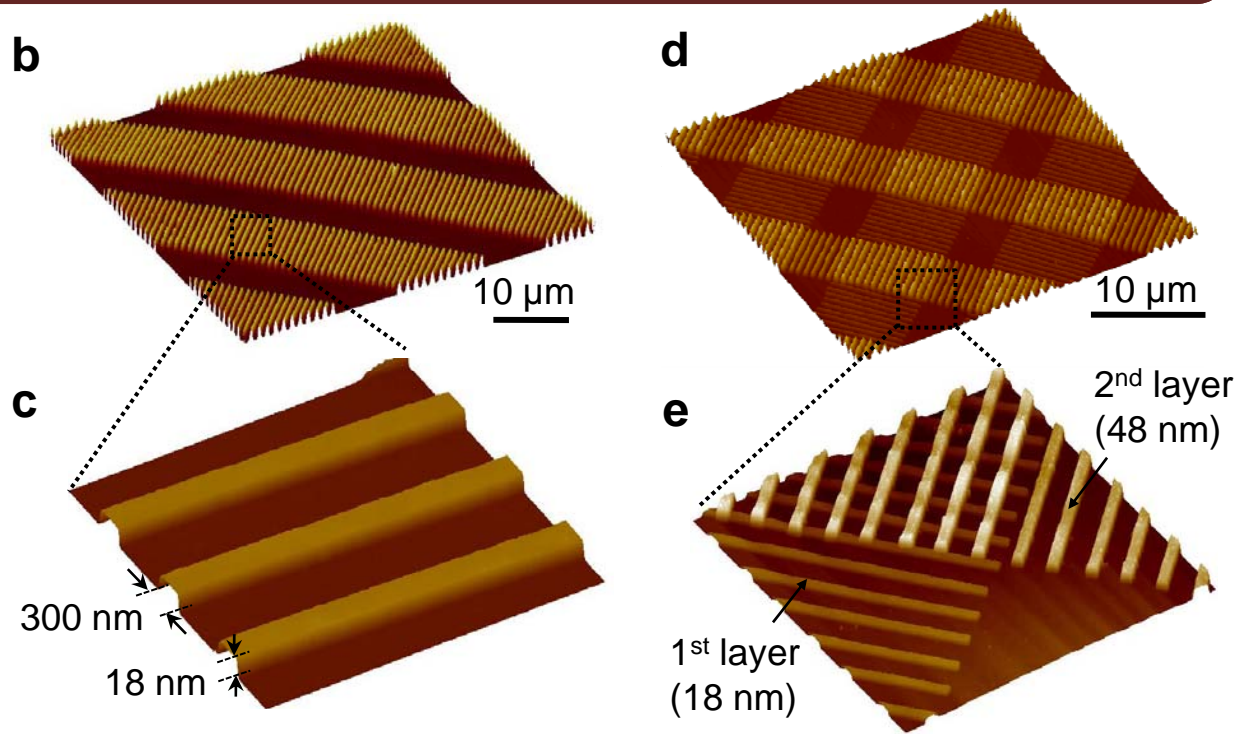

Figure 2

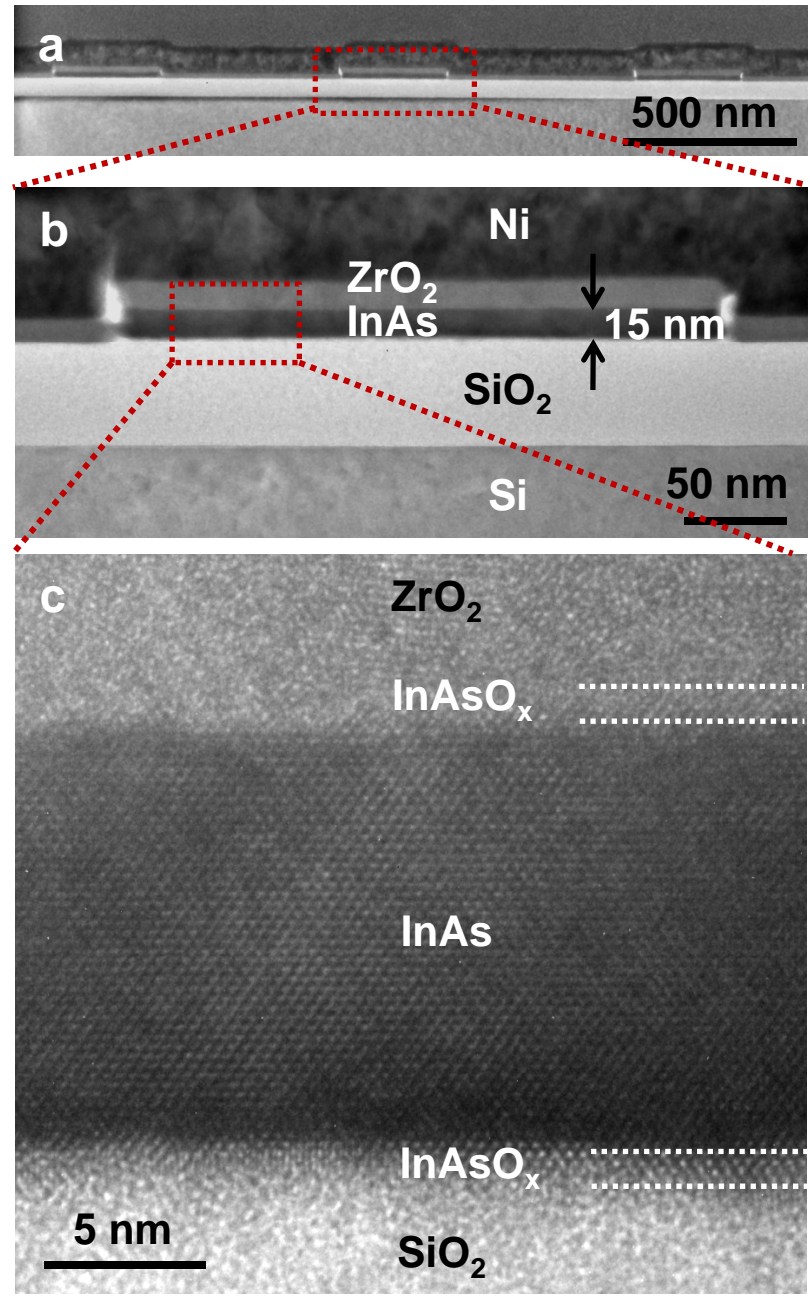

Figure 3

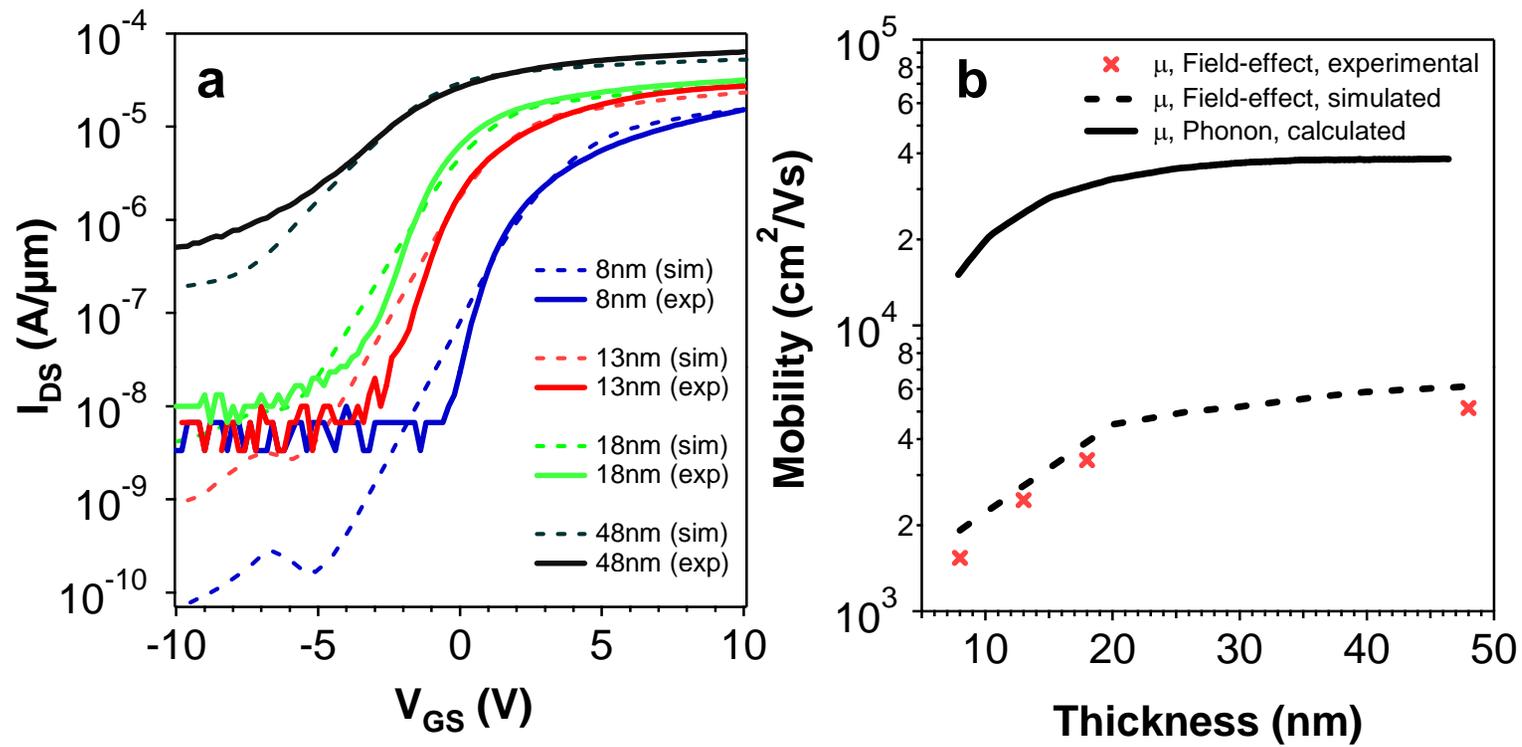

Figure 4

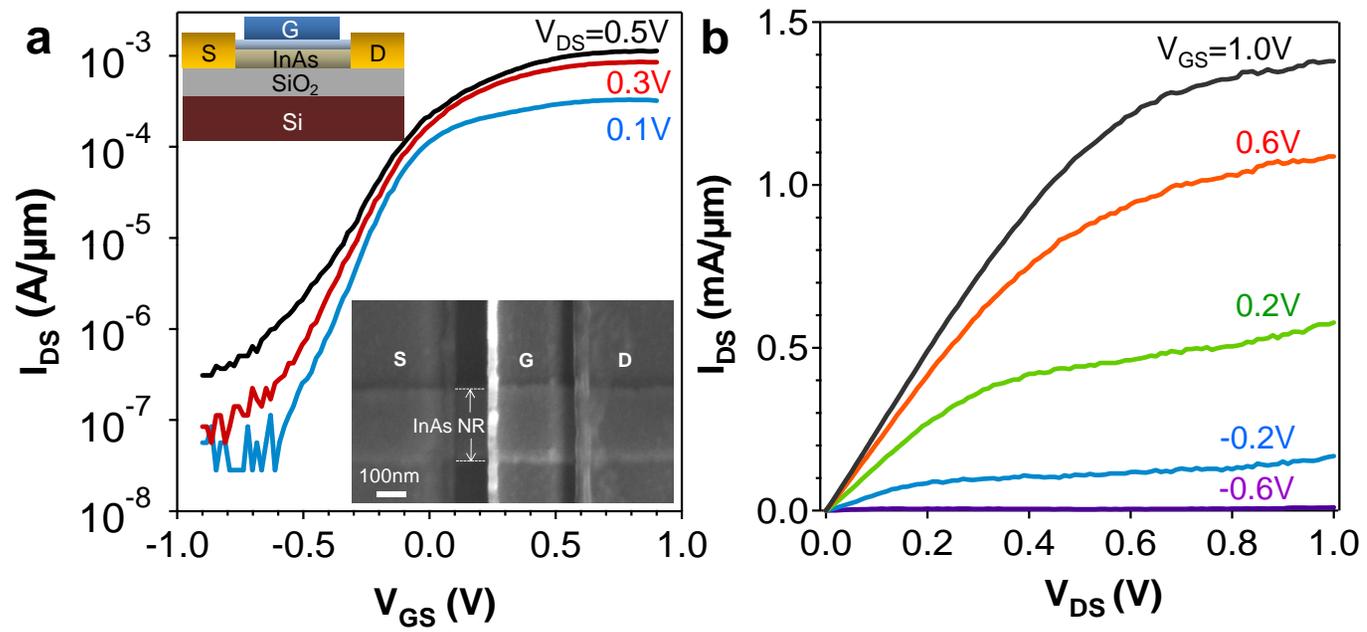

**Ultrathin compound semiconductor on insulator layers for high performance nanoscale transistors**

Hyunhyub Ko[1,2,3,†], Kuniharu Takei[1,2,3,†], Rehan Kapadia[1,2,3,†], Steven Chuang[1,2,3], Hui Fang[1,2,3], Paul W. Leu[1,2,3], Kartik Ganapathi[1], Elena Plis[5], Ha Sul Kim[5], Szu-Ying Chen[4], Morten Madsen[1,2,3], Alexandra C. Ford[1,2,3], Yu-Lun Chueh[4], Sanjay Krishna[5], Sayeef Salahuddin[1], Ali Javey[1,2,3,*]

[1]Electrical Engineering and Computer Sciences, University of California, Berkeley, CA, 94720.
[2]Materials Sciences Division, Lawrence Berkeley National Laboratory, Berkeley, CA 94720.
[3]Berkeley Sensor and Actuator Center, University of California, Berkeley, CA, 94720.
[4]Materials Science and Engineering, National Tsing Hua University, Hsinchu 30013, Taiwan, R. O. C.
[5]Center for High Technology Materials, University of New Mexico, Albuquerque, NM 87106.

[†] These authors contributed equally to this work.
* Correspondence should be addressed to A.J. (ajavey@eecs.berkeley.edu).

# Supplementary Information

**Preparation of the GaSb/Al$_{0.2}$Ga$_{0.8}$Sb/InAs source wafers used for the epitaxial transfer process**

The source layers were grown in a solid source VG-80 molecular beam epitaxy (MBE) reactor on n-type (Te-doped, $5\times10^{17}$ cm$^{-3}$) epi-ready GaSb (001) double-side polished substrates using As$_2$ and Sb$_2$ valved cracker sources. Indium and gallium growth rates were determined by monitoring the intensity oscillations in the reflected high-energy electron diffraction (RHEED) patterns and set to 0.35 ML/s for Ga, 0.30 ML/s for In and 0.43 ML/s for AlGaSb. Group-V fluxes were adjusted using a conventional ion gauge to satisfy a group V/III beam equivalent pressure (BEP) flux ratio equal to 3.6 for GaSb and 9 for InAs. Initially, the substrate was outgassed under a vacuum, and then the surface oxide was removed at high temperature (535 ºC) under an Sb flux. The GaSb and Al$_{0.2}$Ga$_{0.8}$Sb layers of the structure were grown at 490 ºC whereas the InAs layer was grown at 410 ºC. Cross-sectional TEM images of an as-grown source sample is shown Figure S1.

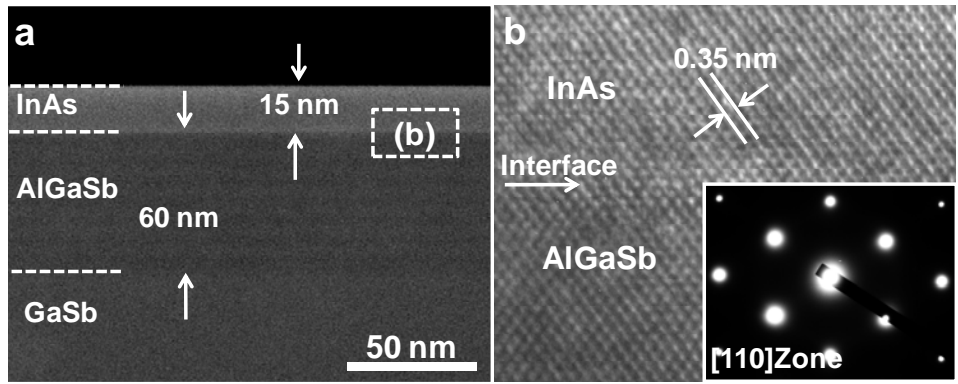

**Figure S1. TEM analysis of the source substrate. a,** The cross-sectional TEM image of a GaSb/AlGaSb/InAs source substrate, showing the InAs thin film (15 nm thick) grown epitaxially on a ~60 nm thick Al$_{0.2}$Ga$_{0.8}$Sb layer on a bulk GaSb wafer. **b,** High-resolution TEM showing the single-crystalline structure of the InAs thin film on AlGaSb. The corresponding diffraction pattern is shown in the inset, indicating the [110] zone.

**Selective wet etching of the AlGaSb sacrificial layer during the epitaxial transfer process**

To release the InAs NRs from the source substrate after the etching of InAs film into NRs, the underlying $Al_{0.2}Ga_{0.8}Sb$ layer was selectively etched. Here, we used ammonium hydroxide (3%, in water) solution for the selective wet etching of the $Al_{0.2}Ga_{0.8}Sb$ layer. Figure S2 shows the SEM images of InAs NRs on the source substrate after different $NH_4OH$ etching times (0, 10, 30, 50 mins), clearly demonstrating the highly selective etching of the AlGaSb sacrificial layer.

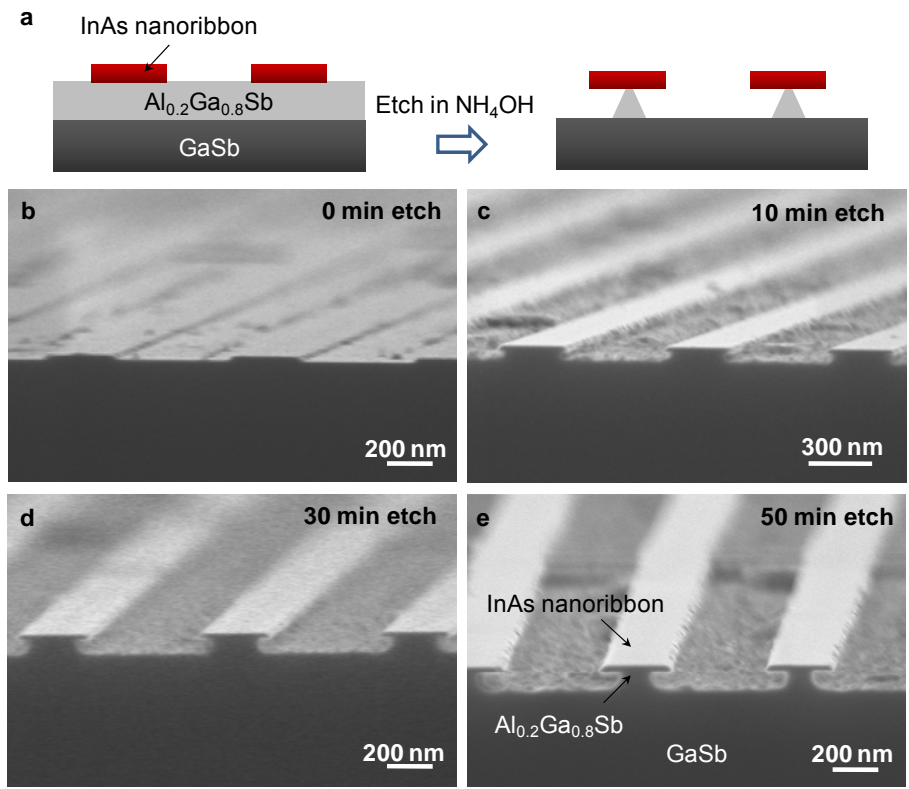

**Figure S2. Selective wet etching of the AlGaSb sacrificial layer. a,** Schematic illustration of the selective etching of AlGaSb. **b-f**, Scanning electron microscopy (SEM) images of InAs NRs on the source substrate after the wet etching of the AlGaSb layer for 0, 10, 30, 50 mins, respectively.

**Details of the epitaxial layer transfer**

Following the partial etch of the AlGaSb layer in ammonium hydroxide (3%, in water) solution for 110 min (for 350 nm wide NRs), an elastomeric polydimethylsiloxane (PDMS) substrate (~2 mm thick) was brought in contact with the source wafer. The PDMS stamp was used to detach the partially released InAs NRs from the source substrate followed by their transfer onto Si/SiO$_2$ (50 nm, thermally grown) receiver substrates. The step by step transfer process is as followed.

1. 10:1 ratio mixture of PDMS prepolymer and curing agent (Sylgard 184, Dow Corning Co., USA) was cured at 80 °C for 4-5 hrs.
2. The cured PDMS slab was cleaned by dipping into toluene for ~30 min and dried completely on top of a hot plate for ~2-3 hrs.
3. The cleaned PDMS slab (~2 mm thick) was pressed (10–200 N/cm$^2$, ~10 sec) on top of the partially released InAs NRs on the source substrate.
4. The PDMS slab was gently detached from the source substrate, resulting in the transfer of InAs NRs from the source wafer to the PDMS slab.
5. The PDMS slab with InAs NRs was dipped in 50:1 HF for 1 min to remove any residues of the sacrificial layer on the surface of InAs NRs.
6. InAs NRs were transferred by pressing (10–200 N/cm$^2$, ~10 sec) the PDMS slab on a Si/SiO$_2$ receiver substrate in ambient laboratory condition (*e.g.* room temperature and air environment). Before the transfer of InAs NRs, the receiver substrate was cleaned by acetone, IPA, and DI water.
7. The PDMS slab was gently removed from the Si/SiO$_2$ substrate, leaving behind the InAs NRs.

Figure S3 shows the SEM images of the source substrate before and after the transfer by a PDMS stamp. The results show that InAs NRs are cleanly cleaved from the source wafer.

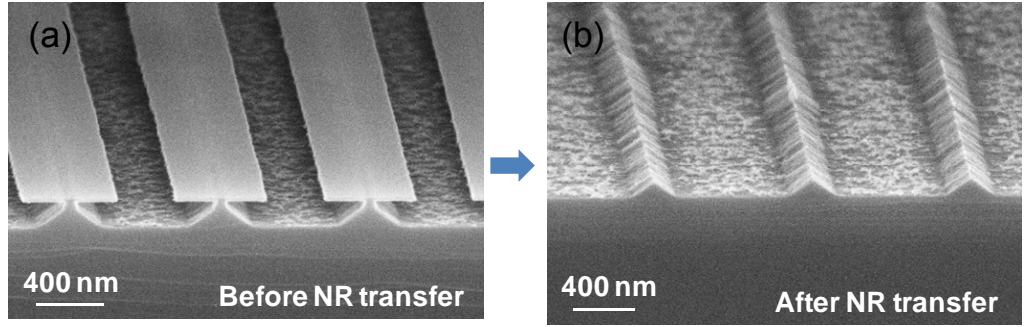

**Figure S3. a,** SEM image of the source substrate after the partial release of InAs NRs (110 min, 3% NH$_4$OH etch) and before the transfer step. **b,** SEM image of the source substrate after the transfer process, showing pyramidal AlGaSb posts.

During the transfer process, residues from the AlGaSb sacrificial layer may remain on the back surface of InAs NRs. To remove any potential residues on the backside, the PDMS slab with InAs NRs was dipped in 50:1 HF for 1 min. From the AFM analysis (Fig. S4), InAs NRs exhibit clean surfaces after the HF treatment, which is critical for making a conformal contact during the subsequent transfer to the Si/SiO$_2$ substrate. To perform AFM analysis of the back side of InAs NRs, a two-step PDMS transfer process was used in which InAs NRs on a PDMS slab were first transferred to a second PDMS slab before getting transferred to a Si/SiO$_2$ substrate. This results in InAs XOI substrates with the original back surface (*i.e.*, on the source wafer) now being on the top. The effect of HF cleaning is clearly depicted in Fig. S4b.

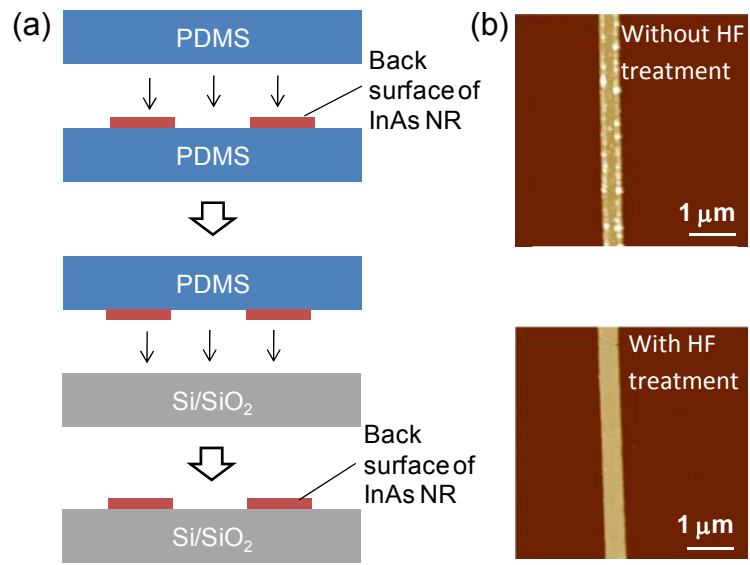

**Figure S4. Effect of HF treatment on the back-surface residues of InAs NRs. a,** Double transfer procedure for the AFM analysis of the back surface of InAs NRs. **b,** AFM images of InAs NRs on a Si/SiO$_2$ substrate without (top) and with (bottom) the use of 50:1 HF treatment for 1 min. The HF treatment was performed while the NRs were on the PDMS slab (*i.e.*, the back surface was exposed), prior to their transfer to the Si/SiO$_2$ substrate.

# Field-effect mobility of long-channel, back-gated XOI FETs based on individual InAs NRs

The transconductance ($g_m = dI_{DS}/dV_{GS}|_{V_{DS}}$) as a function of $V_{GS}$ for back-gated InAs XOI FETs, consisting of individual NRs, was first obtained from the measured transfer characteristics at $V_{DS}$=0.1V. The field-effect electron mobility was then estimated from the relation $\mu_{n,FE} = (g_m)(L^2/C_{ox}V_{DS})$, where $L$ is the channel length and $C_{ox}$ is the gate oxide capacitance. Fig. S5 shows the extracted field-effect electron mobility as a function of $V_{GS}$ for representative XOI FETs with InAs NR thickness of 8, 13, and 48 nm. The peak field-effect mobility increases with the thickness of InAs as depicted in Figure 3b of the main text. It is also evident from the $\mu_{n,FE}$-$V_{GS}$ plots that the field-effect mobility increases with the gate voltage at first and then decreases at high gate voltages due to the enhanced surface scattering of electrons at high electric fields, similar to the conventional MOSFETs.

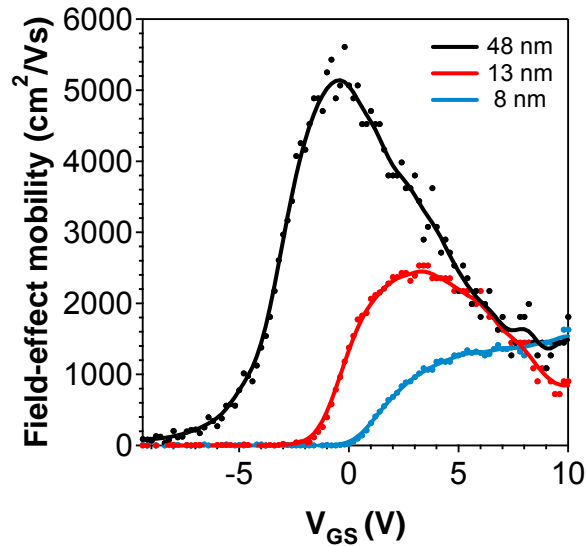

**Figure S5.** Low-field, field-effect mobility of back-gated InAs XOI FETs as a function of $V_{GS}$ for different InAs NR thickness (8, 13, 48 nm) at $V_{DS}$=0.1 V. The field-effect mobility is extracted from the measured $I_{DS}$-$V_{GS}$ curves at $V_{DS}$=0.1 V (Fig. 3a).

**Calculation of phonon mobility of InAs NRs**

This section outlines the calculation of low-field mobility of InAs NRs by considering various phonon scattering mechanisms. As the NRs are not intentionally doped, we assume an electron concentration $n \sim 10^{15}$ cm$^{-3}$ arising due to unintentional doping. The density of states of NRs can be approximated by $m^*/\pi\hbar^2 T_{InAs}$, where $T_{InAs}$ is the NR thickness. Since this density of states is much larger than $n$, it is reasonable to assume that the equilibrium Fermi energy $E_F$ lies within the bandgap for all values of $T_{InAs}$. Hence, the peak mobility measured in experiments corresponds to the maximum transconductance $g_m$, which occurs when $E_F$ coincides with the first conduction sub-band in the channel. We consider the contribution of acoustic and optical phonons along with polar optical phonons - the dominant source of scattering in polar semiconductors like InAs. The scattering rate due to acoustic and optical phonons is summed over longitudinal and transverse modes. The energy dependent scattering rate is averaged over the range of a few $k_B T$ around $E_F$.

$$\left\langle \frac{1}{\tau} \right\rangle = \frac{\int \frac{1}{\tau(E)} \frac{\partial f_0}{\partial E} dE}{\int \frac{\partial f_0}{\partial E} dE}$$

Here $\left\langle 1/\tau \right\rangle$ is the average scattering rate, $1/\tau(E)$ the total scattering rate of an electron with an energy $E$ due to all scattering mechanisms and $f_0$ the equilibrium Fermi-Dirac distribution function. The low-field NR phonon mobility $\mu_{n,phonon}$ is then calculated as $\mu_{n,phonon} = e / m^* \left\langle 1/\tau \right\rangle$, where $e$ is the electronic charge and $m^*$ is the effective mass. An 8×8 Kane's second order $k.p$ Hamiltonian is used to model the quantum confinement effects like the change in the bandgap, effective mass etc. in the dispersion relation of InAs NRs[1,2]. Three approximations are used for the calculations. i) The parabolic band approximation was used for the estimation of the conduction band density of states. This is justified due to the fact that we are interested in the peak mobility that arises at the onset of threshold where the Fermi level is near the bottom edge of the conduction band. For both bulk and thin InAs NRs, the bottom of the conduction band along the direction of transport (100) is largely parabolic. ii) 3D (*i.e.*, bulk) phonon modes were used for all thicknesses[3]. iii) Finally, interband scattering was ignored for simplicity.

The rate for each of the scattering mechanisms is calculated using the Fermi's golden rule wherein the matrix elements of each of the scattering potentials are evaluated in the basis of eigenfunctions of the NR[4].

The scattering rate due to acoustic phonons in a NR of width $T_{InAs}$ is given by[3]:

$$\frac{1}{\tau_{ac}(E)} = \sum_{p=LA,TA1,TA2} \frac{3\pi D_A^2 k_B T}{2 h C_p T_{InAs}} g_{2D}(E)$$

where

$$g_{2D}(E) = \frac{m^*}{\pi \hbar^2} \sum_n \Theta(E - E_n)$$

$$E_n = \frac{h^2}{2m^*_{conf}} \left(\frac{n\pi}{T_{InAs}}\right)^2, \quad n=1,2,3...n_{max}$$

Here, $1/\tau_{ac}(E)$ is the acoustic phonon scattering rate, $D_A$ is the electron intravalley acoustic deformation potential, $C_p$ is the elastic constant corresponding to mode *p*, related to velocity of sound in that mode $v_{s,p}$ by $v_{s,p} = \sqrt{\frac{C_p}{\rho}}$, $\rho$ being the density of InAs, $g_{2D}(E)$ the 2D density of states in the NR, $\Theta(.)$ is the unit step function, $m^*$ is the effective mass in the direction of confinement and $n_{max} = \frac{T_{InAs}}{a_0}$, $a_0$ being the lattice constant of InAs. We used the reported values of $D_A$ and $v_{s,p}$ from Ref. 4 and Ref. 5, respectively.

Similarly, the scattering rate due to optical phonons is given by[3]

$$\frac{1}{\tau_{op}(E)} = \sum_{p=LO,TO} \frac{3\pi D_0^2}{4\rho \omega_p W_{rib}} g_{2D}(E \pm \hbar \omega_p)(N_0 + \tfrac{1}{2} \mp \tfrac{1}{2})$$

where

$$N_0 = \frac{1}{\exp(\frac{\hbar \omega_p}{k_B T}) - 1}$$

Here, $1/\tau_{op}(E)$ is the scattering rate due to optical phonons, $D_0$ the electron optical deformation potential, $\omega_p$ the optical phonon frequency of mode $p$. The top sign corresponds to phonon absorption and bottom one to phonon emission. The values for $\omega_p$ and $d_0$ (=$D_0 a_0$) are obtained from Ref. 6 and Ref. 7, respectively.

The scattering rate due to polar optical phonons is given by[3]:

$$\frac{1}{\tau_{pop}(E)} = \frac{e^2 \omega_{LO}\left(\frac{\kappa_0}{\kappa_\infty} - 1\right)}{2\pi \kappa_0 \varepsilon_0 \hbar \sqrt{2E/m^*}} \left[ N_0 \sinh^{-1}(\tfrac{E}{\hbar \omega_{LO}})^{1/2} + (N_0 + 1)\sinh^{-1}(\tfrac{E}{\hbar \omega_{LO}} - 1)^{1/2} \right]$$

where $1/\tau_{pop}(E)$ is the polar optical phonon scattering rate, $\omega_{LO}$ is the longitudinal optical phonon frequency, $\kappa_0$ and $\kappa_\infty$ are the static and high frequency permitivities respectively. It must be noted that the polar optical phonon scattering rate, owing to the nature of the scattering potential, does not depend explicitly on $T_{InAs}$ unlike the other two scattering mechanisms and the dependence comes through $m^*$.

The calculated mobility vs. thickness is shown in Fig. 3(b) by the solid black curve. For small thicknesses the mobility increases almost linearly with thickness. This is due to the fact that with increasing thickness more modes start to creep into the energy window that contributes to the current flow. As the thickness increases, the additional increase in the number of modes starts to saturate and beyond a threshold point, the mobility saturates to the well known bulk value of InAs[8].

From the measured field-effect and calculated phonon mobilities as a function of $T_{InAs}$ (Fig. 3b), the following observations can be made. First, the calculated value of $\mu_{n,phonon}$ for large values of $T_{InAs}$ (i.e., ~50 nm) is close to the bulk Hall mobility of InAs reported in the literature[8] thus ascertaining that all the dominant scattering mechanisms are considered. Second, the drop in the measured value of field-effect mobility with thickness miniaturization, which signals the onset of confinement effects, occurs for $T_{InAs}$=30-40 nm. This critical thickness which is consistent with the experimental result is close to the Bohr radius of bulk InAs (~34 nm). Notably, the thickness where the system transitions from 3D to 2D depends strongly on $m^*$. A quantitative agreement with experiments in this regard further validates the $m^*$ values calculated from InAs NR dispersion relations. It should be noted that in all the calculations, NRs are effectively treated as thin films, since the widths are large enough (>~300 nm) not to cause confinement effects along the width of the NRs. Only the thickness affects the electronic properties.

**Device simulation of InAs XOI FETs**

The two dimensional simulations were carried out by self consistently solving Poisson's Equation, the electron and hole drift diffusion equations using TCAD Sentaurus 2009. Both top-gated and back-gated device structures were simulated. The back-gated FET consisted of a p-Si substrate with $N_A=10^{21}$ cm$^{-3}$ used as the global gate with 50 nm of SiO$_2$ ($\varepsilon$=3.9) gate dielectric. A 2 nm thick indium oxide layer ($\varepsilon$=3.4) was assumed on the top and bottom surfaces. The channel length was assumed 5 µm, and the InAs thickness was varied from 5-50 nm. The InAs NR was assumed n-type with $N_D=4\times10^{16}$ cm$^{-3}$. This value was chosen to best match the experimental ON current for the devices. In addition, thin regions of heavily doped InAs were inserted between the contacts and the channel to minimize contact effects on the simulated data. Interface traps were placed at the InAs/Indium Oxide interfaces on both the top and bottom surfaces of NRs. The interface trap density was used as a fitting parameter with $D_{it} = 6\times10^{12}$ states eV$^{-1}$-cm$^{-2}$ was found to fit the experimental results the best for all NR thicknesses. The major contribution of $D_{it}$ is reducing the efficiency of the gate-field in modulating the channel charge density. In addition, field-dependent mobility and velocity saturation models were both considered. The interface scattering was treated by using the vertical-field mobility degradation model of Sentaurus. This models the mobility degradation at the interface as a function of the vertical-field using calibrated parameters for conventional MOSFETs. A one-band effective-mass model was used which ignores the effect of quantum confinement on the density of states. In the future, a more accurate device simulation that incorporates the density of states as a function of quantization and InAs thickness is needed. For each NR thickness, the calculated phonon mobility, confined bandgap, and confined effective mass were used as input parameters. Due to the weak gate coupling to the channel (arising from the back-gate geometry) and the high $D_{it}$, the current in the $V_{GS}$ = -0.5V to 0.5V region is not properly handled by Sentaurus. In order to provide for a smooth transition between the subthreshold and accumulation regimes, the simulated $I_{DS}$-$V_{GS}$ curves were fit to an error function, with the points mentioned above removed. This allowed for a more accurate fitting for the region between the subthreshold and ON-state regimes. The threshold voltage of each simulated curve was shifted to match that of the corresponding experimental device. After fitting, the field-effect mobility was deduced as a function of the gate voltage from the simulated $I$-$V$ characteristics by using the analytical expression described previously. The peak mobility was then extracted for each InAs thickness and plotted in Figure 3b. Note that the 5-10x reduction of the field-effect mobility as compared to the calculated phonon mobility is expected due to the various device contributions, including vertical-field induced carrier scattering and $D_{it}$. As a parallel, for lightly doped Si, the measured mobility for an effective field of 1MV/cm is ~250 cm$^2$/V-s, while, for the same doping, the sheet mobility is ~1100 cm$^2$/V-s [9,10]. This behavior is similar to that observed in our InAs XOI FETs.

Similarly, the top-gated XOI FETs were simulated with 2 nm of indium oxide assumed on the two surfaces of InAs with a body doping concentration of $N_D=4\times10^{16}$ cm$^{-3}$. The top gate stack was composed of 7 nm of ZrO$_2$ ($\varepsilon$=20) and a metal gate electrode with a workfunction of 5eV. The source and drain contacts were assumed ohmic. To fit the subthreshold swing of the experimental devices, the trap density at the InAs/InAsO$_x$ interfaces was chosen to be $D_{it}=10^{11}$ states eV$^{-1}$-cm$^{-2}$. Notably, this extracted $D_{it}$ is ~60× lower than that of the back-gated FETs as the former consists of thermally grown InAsO$_x$ passivation layer while the latter consists of a native oxide layer. To fit the linear region, the series resistance at the source and drain ($R_s$, $R_d$) were chosen to be 100 Ω.μm (unit width normalized) as obtained from the length dependent transport studies (Fig. S6). The threshold voltage was shifted to match that of the corresponding experimental device. The simulation results are show in Figure S7, clearly depicting the close match between the experiment and simulation, further demonstrating the near ideal material and device system presented in this work with deterministic electrical properties. Note that this simulation is valid only for low to moderate gate overdrives where the Fermi level is near the conduction band edge. Within this regime, the parabolic band approximation used in this work is valid. In the future, a more accurate simulation of the XOI FETs with the proper band structure treatment and quantum confinement effects is needed.

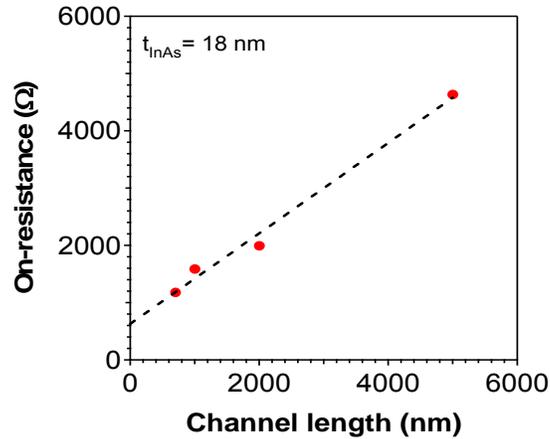

**Figure S6. Extraction of the parasitic contact resistance.** ON-state resistance vs. channel length, $L$ for single NR FETs (back-gated) with $T_{InAs}$~18 nm and width ~350 nm. On-resistances are extracted from back-gated $I_{DS}$-$V_{GS}$ curves at $V_{GS}$=10V and $V_{DS}$=0.3V. The experimental data are shown as red dots and the dashed line is the best fit line. The extrapolated resistance at $L$=0 nm corresponds to the parasitic contact resistance. From the data, a parasitic resistance of ~600 $\Omega$ is obtained, corresponding to ~300 $\Omega$ for the S/D electrodes individually. The unit width normalized parasitic resistance for each contact electrode is ~100 $\Omega.\mu m$.

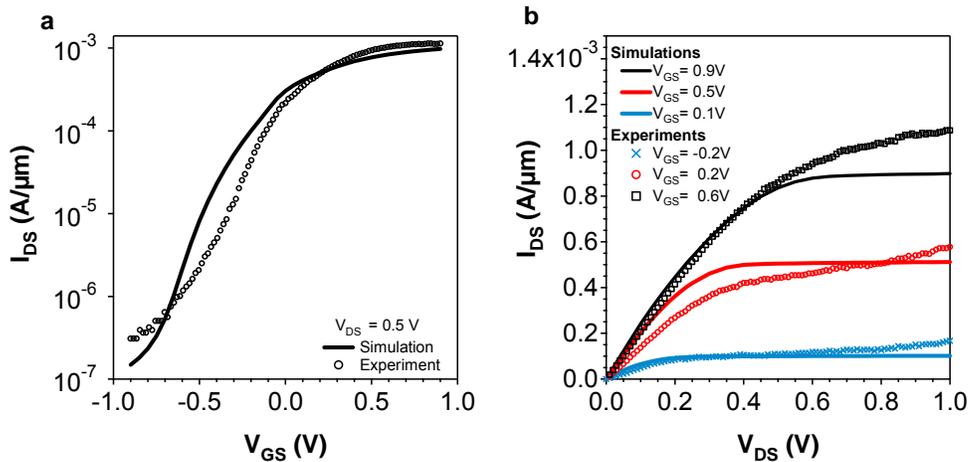

**Figure S7. Electrical characterization of top-gated InAs XOI FETs. a**, Transfer and **b**, output characteristics of an InAs XOI FET (~18 nm thick) with $L$~0.5 µm, showing a close fit between the experiment and simulation. Note that the device is the same as the one shown in Figure 4b-c of the main text.

**Electrical properties of InAs XOI top-gated FETs as a function of surface/interface treatment**

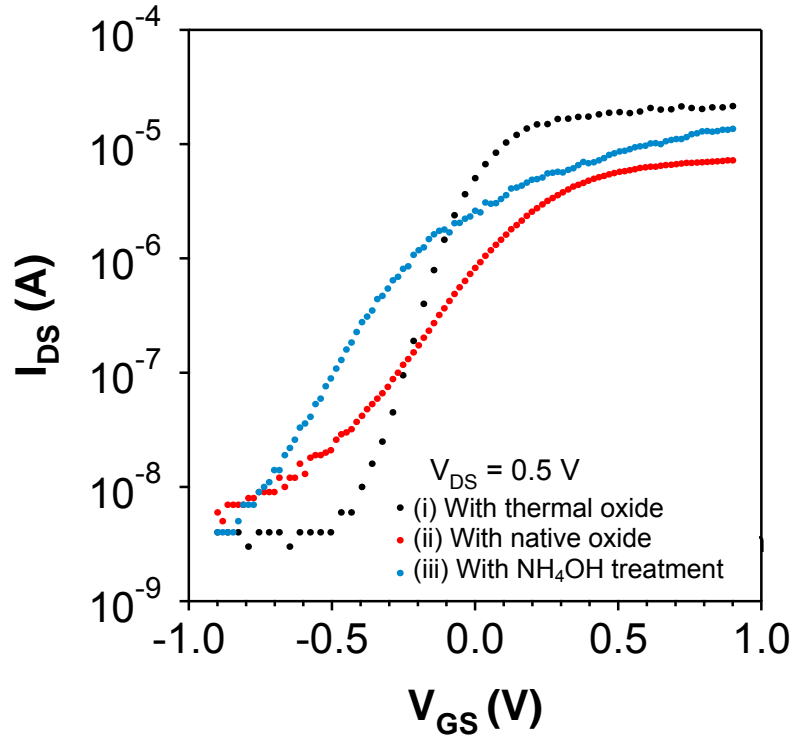

**Figure S8**. **Transfer characteristics of three InAs XOI top-gated FETs with different surface treatment prior to the ALD of the ZrO$_2$ gate dielectric.** (i) With a thermal oxidation of InAs at 350ºC for 1 min (resulting in ~1 nm thermal InAsO$_x$) prior to the ALD (black marks), (ii) without any surface treatment (*i.e.*, consisting of ~ 1 nm native surface oxide layer) before ALD (red marks), and (iii) with NH$_4$OH immediately prior to the ALD to remove the surface oxide layer (blue marks). For the thermally oxidized sample, the native oxide was first removed by a treatment with 3% NH$_4$OH. The results clearly depict the drastic enhancement of the subthreshold characteristics due to the effective surface passivation role of the thermally grown InAsO$_x$ layer, resulting in enhanced electrostatic coupling of the gate electrode. The SS is 107, 290, and 230 mV/decade for devices (i)-(iii), respectively. The channel lengths are 2 μm, 5 μm, and 5 μm for devices (i)-(iii), respectively.

**Experimental transconductance as a function of gate bias for a top-gated XOI FET**

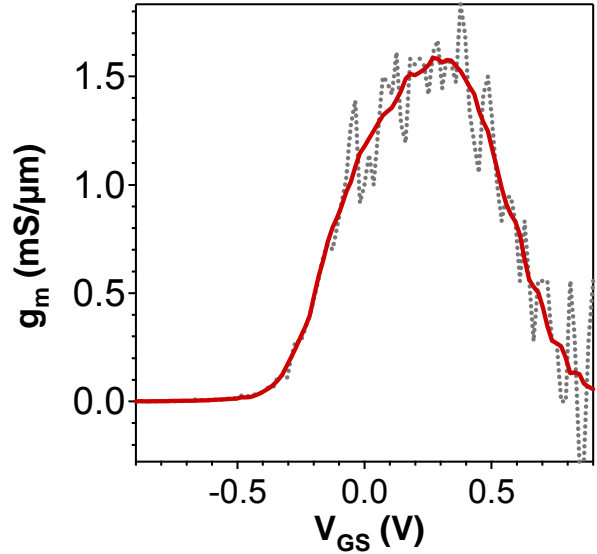

**Figure S9. Transconductance, $g_m = dI_{DS}/dV_{GS}|_{V_{DS}}$ at $V_{DS}=0.5$ as a function of $V_{GS}$ obtained from the $I_{DS}$-$V_{GS}$ data shown in Figure 4c.** The dashed gray line represents the obtained transconductance after current differentiation while the red line is after 2$^{nd}$ order Savitsky-Golay smoothing.

The extrinsic transconductance, $g_m$, extracted from the measurement (Fig. S9) includes the effect of series resistance. The intrinsic transconductance, $g_{mi}$, can be extracted as $g_{mi} = g_m/(1-g_m R_S - g_d R_{SD})$[11], where $R_S$ is the source series resistance, $R_{SD}$ is the source and drain series resistance (i.e., $R_{SD}=R_S+R_D$), and $g_d$ ($=dI_{DS}/dV_{DS}$) is the drain conductance. Using this analysis, the peak $g_{mi}$ of the experimental device is ~2 mS/μm, assuming $R_S=R_D=100$ Ω.μm (Fig. S6) and $g_d \sim 0.23$ mS/μm (at $V_{GS}=0.2$V and $V_{DS}=0.5$V; Fig. 4b).

To analyze the transconductance, it is beneficial to look at the basic equation for drain current in a MOSFET, $I_{DS} = v_{drift} \times n \times q$, where $v_{drift}$ is the electron drift velocity, n is the electron density, and q is the charge of an electron. At low electric-fields, $v_{dirft}=\mu_n \times E$, where E is the electric field while at high fields, the velocity saturates at $v_{sat}$. The electron density, n can be approximated as $n=(C_{ox}/L)\times(V_{GS}-V_t)$, where $V_t$ is the threshold voltage and $C_{ox}$ is total gate capacitance. For long channel lengths and low electric-fields, $g_{mi}=\mu_n C_{ox}(V_{GS}-V_t)/L^2$ while $g_{mi}=v_{sat}(C_{ox}/L)$ in the high-field regime. Based on the velocity vs. field for bulk InAs from literature[12], our experimental devices, with L~0.5μm and $V_{ds}$=0.5 V, are operating closer to the high-field limit rather than the diffusive regime. Since $C_{ox}$ is the total capacitance, it is a linear function of the channel length, L, making $g_{mi}$ independent of L at high-fields. From the $v_{drift}$ vs. electric-field curve[12] for bulk InAs, at a field of 10 kV/cm (corresponding to $V_{DS}$=0.5V and L=0.5 μm used in the experiments), the drift velocity is ~1.5 x $10^7$ cm/s. For the presented device with $C_{ox}$~2×$10^{-15}$ F ($t_{ox}$~8 nm $ZrO_2$, L=0.5 μm and width, W=320 nm), $g_{mi}$ ~ 0.61 mS is approximated by using the high-field analytical expression, corresponding to a unit-width normalized value of ~2 mS/μm. This calculated value is close to that of the experimentally extracted value. Note that this presents only a rough guideline since the bulk velocity vs. field curve was used in the calculation as that of an ultrathin InAs layer is not well established in the literature. When compared to InAs HEMTs in the literature, the transconductance of our FETs is comparable. The value of intrinsic transconductance for InAs quantum-well FETs (QWFETs) with 10 nm InAs thickness is reported to be ~3mS/ μm for L=40 - 200 nm, with minimal dependence on L for this explored range[13]. This is expected since for such short channel lengths, the devices are operating in the high-field regime where the intrinsic transconductance shows minimal length dependence as noted above.

At high gate fields (i.e., $V_{GS}$>0.3V), the transconductance rapidly drops. This can be attributed to a combination of various factors which effectively cause a current saturation at high gate-fields. These effects include (i) an increase in the electron effective mass and thereby a reduction in the carrier velocity as the Fermi level is driven deep into the conduction band due to the band structure of InAs NRs, (ii) enhanced surface scattering of carriers at high vertical fields, similar to that of the conventional MOSFETs, and (iii) the inability of the metal source to supply enough charge due to the metal-semiconductor contacts as shown previously for thin-body, metal-contacted MOSFETs[14]. In the future, a more in-depth exploration of transport physics of careers is required to better elucidate the observed field-dependent behavior.

**InAs XOI FET with the lowest observed subthreshold swing**

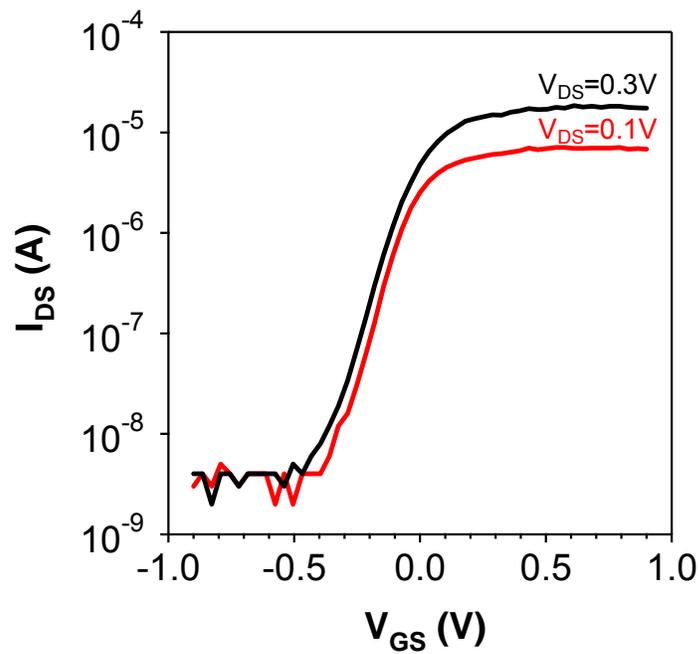

**Figure S10**. **The $I_{DS}$-$V_{GS}$ characteristics of a top-gated InAs XOI FET, consisting of an individual NR with ~18 nm thickness.** The channel length is ~2 μm, gate dielectric ($ZrO_2$) thickness is ~6 nm deposited by ALD. The subthreshold swing, SS~107mV/dec.

**Hysteresis measurements of top-gated XOI FETs**

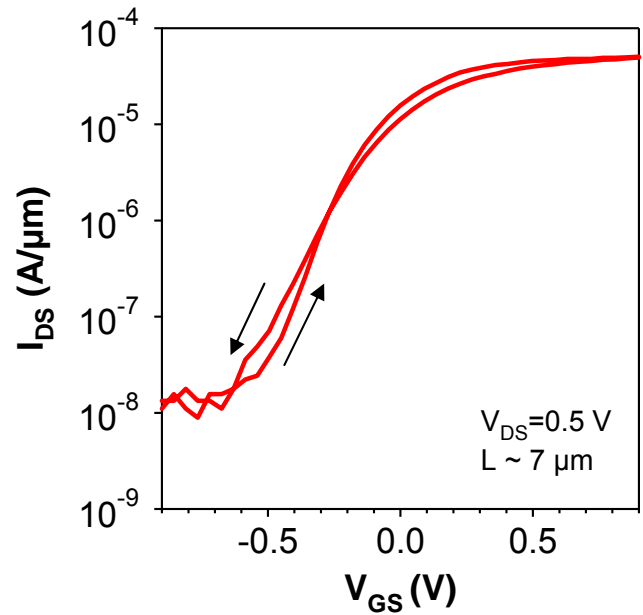

**Figure S11.** Backward and forward sweep $I_{DS}$-$V_{GS}$ curves for a top-gated XOI FET with ~7 μm gate-length and $V_{DS}$=0.5 V. The hysteresis magnitude is relatively small given that the XOI processing (i.e., NR transfer) steps were performed outside of a cleanroom environment.

**InAs XOI device variation**

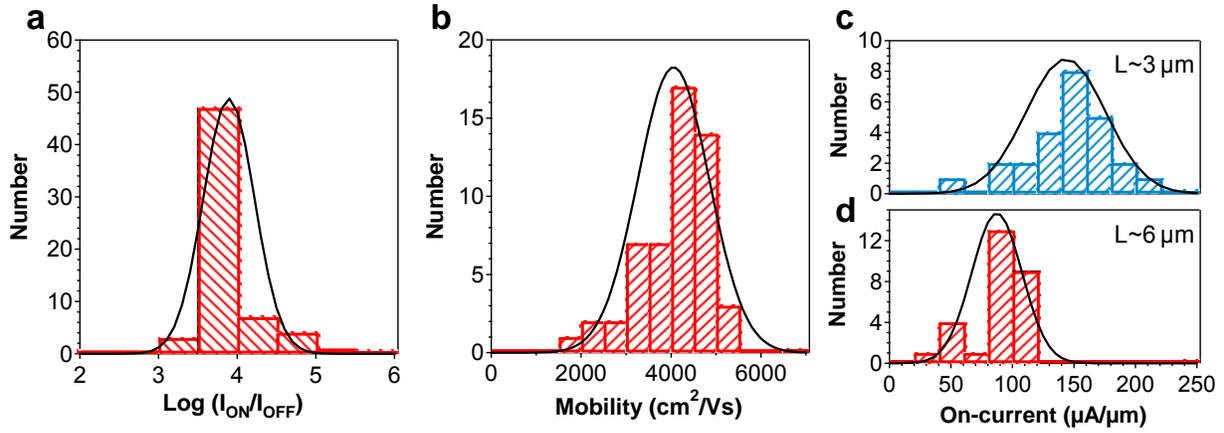

**Figure S12.** Histogram plots of **a**, $I_{ON}/I_{OFF}$ ratio, **b,** field-effect mobility, and **c,** $I_{ON}$ (normalized by the width of NRs) for InAs XOI FETs. The parameters of the back-gated ($T_{SiO2}$ ~ 50 nm) devices were as follows: $T_{InAs}$~13 nm, $L_G$~3 and 6 μm, and $V_{DS}$ = 0.3 V.

**InAs *micro*-ribbon FETs**

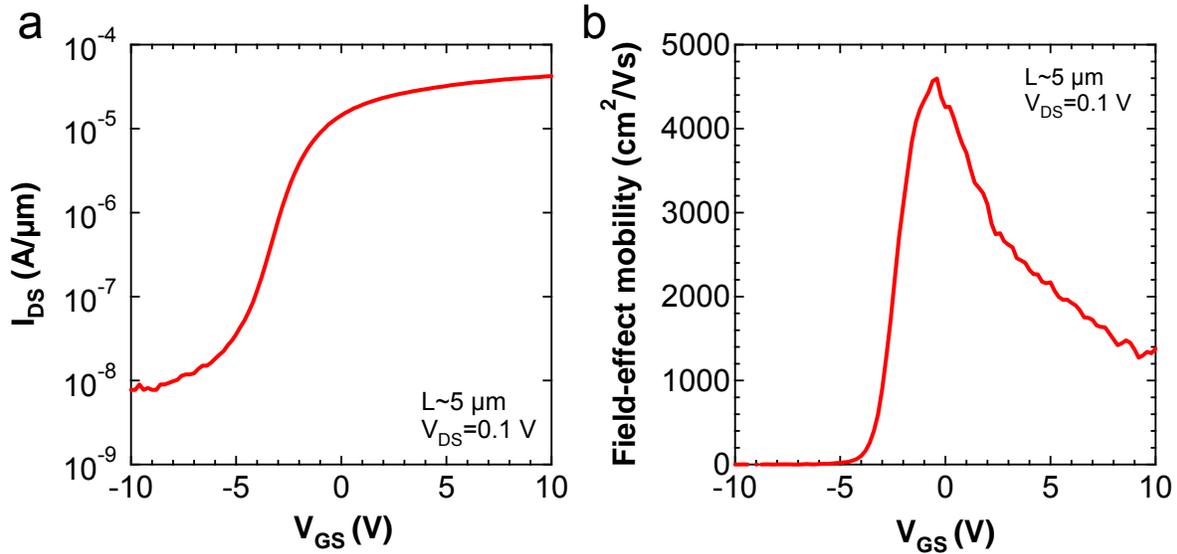

**Figure S13. A back-gated XOI FET based on a single 2.5 μm wide, 18nm thick InAs ribbon. a,** $I_{DS}$-$V_{GS}$ curve, and **b,** field-effect mobility as a function of the gate-field. The characteristics are similar to the devices made from 350nm wide InAs NRs, providing evidence that the width of the ribbons does not play a role in the device performance, as long as they are wider than a few hundred nm (i.e., width >>thickness) which is the case in this work.

**References**

1. Gershoni, D., Henry, C. H., Baraff, G. A. Calculating the optical properties of multidimensional heterostructures: application to modeling of quaternary quantum well lasers. *J. Quantum Electron*. **29**, 2433–2450 (1993).

2. Foreman, B. A. Elimination of spurious solutions from eight-band k. p theory. *Phys. Rev. B* **56**, R12748 (1997).

3. Mark Lundstrom. Carrier scattering. In *Fundamentals of Carrier Transport* (Cambridge, 2000).

4. Van de Walle, C. G. Band lineups and deformation potentials in the model-solid theory. *Phys. Rev. B* **39,** 1871-1883 (1989).

5. Adachi, S. Indium arsenide (InAs). *Handbook on Physical Properties of Semiconductors* (Kluwer Academic Publishers, 2004).

6. Groenen, J., Priester, C., Carles, R. Strain distribution and optical phonons in InP/InAs self-assembled quantum dots. *Phys. Rev. B* **60,** 16013-16017 (1999).

7. Pötz, W., Vogl, P. Theory of optical-phonon deformation potentials in tetrahedral semiconductors. *Phys. Rev. B* **24**, 2025 – 2037 (1981).

8. Mikhailova, M. P. Indium arsenide. *Handbook Series of Semiconductor Parameters, vol 1: Elementary Semiconductors and A3B5 Compounds Si, Ge C, GaAs, GaP, GaSb InAs, InP, InSb* (World Scientific, 1996).

9. Chau, R. et al., Advanced CMOS Transistors in the Nanotechnology Era for High-Performance, Low-Power Logic Applications, *ICSICT Tech. Dig.* 26-30 (2004).

10. Sze, S. M., Irvin, J. C. Resistivity, Mobility and Impurity Levels in GaAs, Ge, and Si at 300K. *Solid-State Electronics* **11**, 599-602 (1968).

11. Chou, S. Y., Antoniadis, D.A., Relationship Between Measured and Intrinsic Transconductances of FET's, *IEEE Trans. Electron Devices* **34**, 448 – 450 (1987).

12. Brennan, K., Hess, K., High Field Transport in GaAs, InP and InAs, *Solid-State Electronics* **27**, 347-357, (1984).

13. Kim, T.-W., Kim, D.-H., del Alamo, J. A., Logic characteristics of 40 nm thin-channel InAs HEMTs, *Indium Phosphide and Related Materials (IPRM),* 2010.

14. Guo, J., Lundstrom M. S., A computational study of thin-body, double-gate, Schottky barrier MOSFETs, *IEEE Trans. Electron Devices* **49**, 1897-1902 (2002).